\documentclass[preprint2]{aastex6}
\usepackage{amsfonts,amsmath,amssymb,ulem,nicefrac,verbatim,color}
\usepackage{psfig,array,graphicx,longtable,affils}

\def\simlt{\mathrel{\hbox{\rlap{\hbox{\lower4pt\hbox{$\sim$}}}\hbox{$<$}}}}
\def\simgt{\mathrel{\hbox{\rlap{\hbox{\lower4pt\hbox{$\sim$}}}\hbox{$>$}}}}

\def\ale{\mathrel{\hbox{\rlap{\hbox{\lower4pt\hbox{$\sim$}}}\hbox{$<$}}}}
\def\age{\mathrel{\hbox{\rlap{\hbox{\lower4pt\hbox{$\sim$}}}\hbox{$>$}}}}

\def\spose#1{\hbox to 0pt{#1\hss}}

\begin{document}

\title{Radio observations of the tidal disruption event XMMSL1 J0740$-$85}

\DeclareAffil{cfa}{Harvard-Smithsonian Center for Astrophysics, 60 Garden St., Cambridge, MA 02138, USA}
\DeclareAffil{atnf}{Australia Telescope National Facility, CSIRO Astronomy and Space Science, PO box 76, Epping, NSW 1710, Australia}
\DeclareAffil{xmm}{XMM SOC, ESAC, Apartado 78, 28691 Villanueva de la Ca\~nada, Madrid, Spain}
\DeclareAffil{china}{QianNan Normal University for Nationalities, Longshan Street, Duyun City of Guizhou Province, China}

\affilauthorlist{K.~D.~Alexander\affils{cfa}, M.~H.~Wieringa\affils{atnf}, E.~Berger\affils{cfa}, R.~D.~Saxton\affils{xmm}, S.~Komossa\affils{china}}

\begin{abstract}
We present radio observations of the tidal disruption event candidate (TDE) XMMSL1 J0740$-$85 spanning 592 to 875 d post X-ray discovery.  We detect radio emission that fades from an initial peak flux density at 1.6 GHz of $1.19\pm 0.06$ mJy to $0.65\pm 0.06$ mJy suggesting an association with the TDE. This makes XMMSL1 J0740$-$85 at $d=75$ Mpc the nearest TDE with detected radio emission to date and only the fifth TDE with radio emission overall.  The observed radio luminosity rules out a powerful relativistic jet like that seen in the relativistic TDE Swift J1644+57.  Instead we infer from an equipartition analysis that the radio emission most likely arises from a non-relativistic outflow similar to that seen in the nearby TDE ASASSN-14li, with a velocity of about $10^4$ km s$^{-1}$ and a kinetic energy of about $10^{48}$ erg, expanding into a medium with a density of about $10^2$ cm$^{-3}$.  Alternatively, the radio emission could arise from a weak initially-relativistic but decelerated jet with an energy of $\sim 2\times 10^{50}$ erg, or (for an extreme disruption geometry) from the unbound debris.  The radio data for XMMSL1 J0740$-$85 continues to support the previous suggestion of a bimodal distribution of common non-relativistic isotropic outflows and rare relativistic jets in TDEs (in analogy with the relation between Type Ib/c supernovae and long-duration gamma-ray bursts).  The radio data also provide a new measurement of the circumnuclear density on a sub-parsec scale around an extragalactic supermassive black hole.
\smallskip
\end{abstract}


\section{Introduction}

In recent decades bright flares in the nuclei of several dozen previously-quiescent galaxies have been interpreted as transient accretion onto supermassive black holes (SMBHs) caused by the tidal disruption of a star \citep{rees88, kom15}. The primary predicted observational signature of these tidal disruption events (TDEs) is transient thermal emission from the newly-formed accretion disk, peaking at extreme ultraviolet (UV) wavelengths. Detailed multi-wavelength follow-up of TDE candidates in recent years has revealed soft X-rays, UV, and optical emission that point to a more complicated picture, including likely reprocessing of the disk emission by outflows (recent review by \citealt{kom15}). Additionally, three TDEs have been discovered to launch relativistic jets, detected on-axis in $\gamma$-rays, hard X-rays, and in two cases radio (e.g. \citealt{bgm+11,bkg+11,ltc+11,zbs+11,ckh+12,bls+15}). {\it Swift} J164449.3+573451 (hereafter Sw J1644+57) is the prototypical jetted TDE and is still observable in the radio band more than five years after discovery. Observations of Sw J1644+57 have enabled new insights into the formation, evolution, and cessation of relativistic jets from SMBHs and have provided the first picture of the circumnuclear density profile of a quiescent $z=0.354$ galaxy on sub-parsec scales \citep{zbs+11,bzp+12,zbm+13}. Radio observations of TDEs also provide an independent measurement of the event energy, the size of the emitting region, and the magnetic field strength (e.g. \citealt{zbs+11,bzp+12,zbm+13,abg+16,vas+16,lei16}). 

We expect mass ejection and therefore radio emission due to interaction with circumnuclear matter for most, if not all TDEs, as theoretical models predict that the initial fallback rate for most events should be super-Eddington \citep{sq09, vkf11, gm11, gr13}. However, only four TDEs with associated radio emission have been published to date: two jetted events discovered by {\it Swift} (Sw J1644+57 and Sw J2058+0516), IGR J1258+0134, claimed to have an off-axis relativistic jet \citep{ir15,lei16}, and ASASSN-14li, which produced less luminous radio emission arising from a non-relativistic outflow \citep{abg+16,vas+16}. Radio upper limits for an additional 15 events rule out Sw J1644+57-like jets in most cases, but cannot rule out slower, non-relativistic outflows as seen in ASASSN-14li \citep{kom02,bmc+13,vfk+13,cbg+14,ags+14}. Building on this effort, we have begun a systematic effort to obtain radio observations of nearby TDE candidates, for which even non-relativistic outflows should be detectable with current facilities. 

On 2014 April 1 UT, the XMM-{\it Newton} X-ray satellite detected a flare from the nucleus of the nearby ($z=0.0173$; $d=75$ Mpc) quiescent galaxy 2MASX 07400785$-$8539307 as part of the XMM-{\it Newton} slew survey \citep{sax08}. The flare (hereafter XMMSL1 J0740$-$85) was discovered to extend from the hard X-ray band through the UV, with minimal variability in the optical, and consists of both thermal and non-thermal components \citep{sax16}. It reached a peak bolometric luminosity of $\sim2\times10^{44}$ erg s$^{-1}$ before decreasing by a factor of 70 in the X-rays and 12 in the UV over $\sim530$ d and was interpreted by \cite{sax16} as a TDE. The X-ray variability constrains the SMBH mass to be $M_{\rm BH}\approx 3.5\times10^6$ M$_\odot$, consistent with this interpretation \citep{sax16}. The host galaxy exhibits no current star formation or AGN activity, and its optical spectrum is consistent with a burst of star formation $\sim2$ Gyr ago, placing it within the rare category of post-starburst galaxies seemingly favored by recent TDE candidates \citep{ags+14,faz16}. Motivated by an exploratory radio detection consistent with the nucleus of the host galaxy \citep{sax16}, we undertook a radio monitoring campaign of XMMSL1 J0740$-$85 to determine if the radio emission is associated with the TDE. Here we present the results and analysis of this campaign.

This paper is structured as follows. In Section \ref{sec:obs}, we present our radio observations of XMMSL1 J0740$-$85, spanning $592-875$ d after discovery. In Section \ref{sec:mod}, we outline possible models for the radio emission. We then use a Markov Chain Monte Carlo (MCMC) analysis to constrain the physical properties of the outflow launched by the TDE, as well as the circumnuclear density. We compare these results to those obtained for other TDEs with radio emission in Section \ref{sec:disc}, and present our conclusions in Section \ref{sec:conc}.

\section{Radio Observations}\label{sec:obs}

\begin{figure*}
\begin{center}
\includegraphics[width=6.4in]{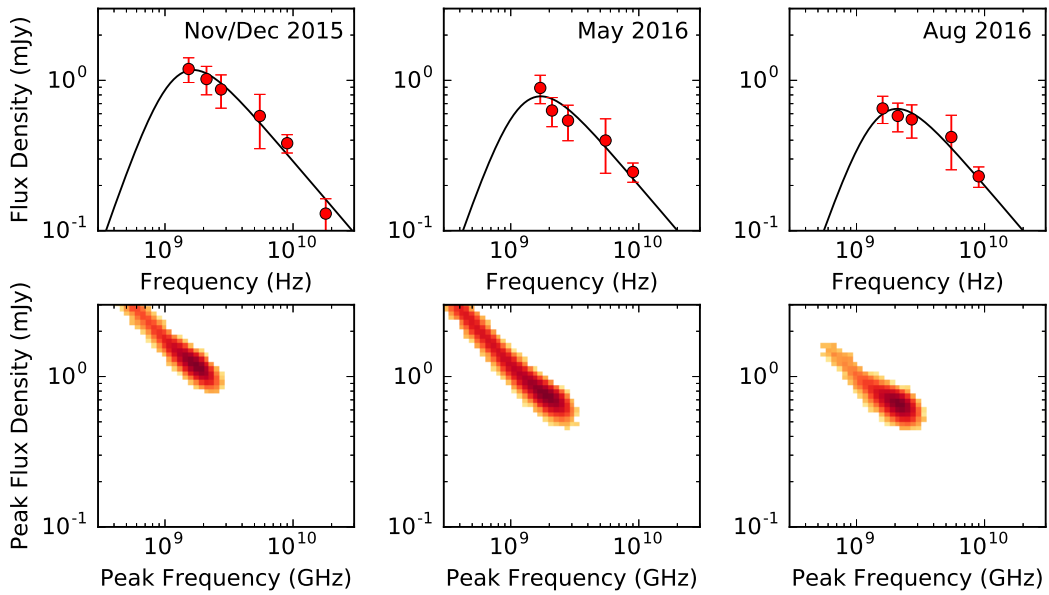}
\end{center}
\caption{Radio data for XMMSL1 J$0740-85$ (red circles) along with the results of our MCMC modeling of the radio emission (black lines). The errorbars include statistical, calibration, and scintillation-induced uncertainties. The second row shows a two-dimensional histogram of the MCMC output for each epoch.} 
\label{fig:mcmc}
\end{figure*}

\begin{table}
\setlength\LTcapwidth{2.5in}
\caption{\label{tab:obs}Radio observations of XMMSL1 J0740$-$85}
\begin{tabular}{lcccc}
\hline
\hline\noalign{\smallskip}
UT Date & $\Delta t$ & $\nu$ & $F_\nu$ $\pm$ stat $\pm$ ISS & Config- \\
             & (days)         & (GHz)  & (mJy) & uration\\ 
\hline\noalign{\smallskip}
2015 Nov 14 & 592 & 5.5 & 0.58 $\pm$ 0.01 $\pm$ 0.22 & 6A \\
2015 Nov 14 & 592 & 9.0 & 0.38 $\pm$ 0.01 $\pm$ 0.04 & 6A \\
\hline\noalign{\smallskip}
2015 Dec 1 & 609 & 1.5 & 1.19 $\pm$ 0.06 $\pm$ 0.18  & 1.5A \\ 
2015 Dec 1 & 609 & 2.1 & 1.02 $\pm$ 0.04 $\pm$ 0.19  & 1.5A \\ 
2015 Dec 1 & 609 & 2.7 & 0.87 $\pm$ 0.04 $\pm$ 0.19 & 1.5A \\ 
2015 Dec 1 & 609 & 18.0 & 0.13 $\pm$ 0.03 $\pm$ 0.005 & 1.5A \\ 
\hline\noalign{\smallskip}
2016 May 9 & 769 & 1.7 & 0.89 $\pm$ 0.09 $\pm$ 0.14 & 6A \\ 
2016 May 9 & 769 & 2.1 & 0.63 $\pm$ 0.04 $\pm$ 0.12 & 6A \\ 
2016 May 9 & 769 & 2.8 & 0.54 $\pm$ 0.05 $\pm$ 0.12 & 6A \\ 
2016 May 9 & 769 & 5.5 & 0.40 $\pm$ 0.01 $\pm$ 0.15 & 6A \\ 
2016 May 9 & 769 & 9.0 & 0.25 $\pm$ 0.01 $\pm$ 0.02 & 6A \\ 
\hline\noalign{\smallskip}
2016 Aug 23 & 875 & 1.6 & 0.65 $\pm$ 0.06 $\pm$ 0.10 & 6C \\ 
2016 Aug 23 & 875 & 2.1 & 0.58 $\pm$ 0.03 $\pm$ 0.11 & 6C \\
2016 Aug 23 & 875 & 2.7 & 0.55 $\pm$ 0.03 $\pm$ 0.12 & 6C \\
2016 Aug 23 & 875 & 5.5 & 0.42 $\pm$ 0.02 $\pm$ 0.16 & 6C \\
2016 Aug 23 & 875 & 9.0 & 0.23 $\pm$ 0.02 $\pm$ 0.02 & 6C \\
\hline\noalign{\smallskip}
\end{tabular}
{\bf Notes.} All values of $\Delta t$ are relative to 2014 April 1 UT, the discovery date in X-rays. The flux values are given with associated statistical uncertainties from fitting a point source model to the imaged data and the additional flux variation expected from interstellar scintillation (ISS). The ATCA telescope configuration is given in the rightmost column. Our December 2015 observation only used five antennas, as CA03 was unavailable.  
\end{table}

We observed the position of XMMSL1 J0740$-$85 with the Australia Telescope Compact Array (ATCA) beginning on 2015 November 14 UT, 592 d after the initial X-ray discovery. In our initial observation, we detected a source at $\alpha= 07^{\text h}40^{\text m}08\fs$19, $\delta=-85\degr39\arcmin31\farcs25$ ($\pm 0\farcs3$ in each coordinate) at 5.5 GHz and 9.0 GHz. This is consistent with the {\it Swift} UVOT position ($\alpha= 07^{\text h}40^{\text m}08\fs$43, $\delta=-85\degr39\arcmin31\farcs4$, 90\% confidence radius $0\farcs4$), the X-ray position, and the nucleus of the host galaxy \citep{sax16}. Further observations on 2015 December 1 UT resulted in additional detections at 2.1 GHz and 18 GHz. We observed the source twice more under program C3106 on 2016 May 9 UT and 2016 August 23 UT (see Table \ref{tab:obs}).

We analyzed the data using the Miriad package \citep{stw95}. The data were flagged for RFI and calibrated using PKSB1934$-$638 as the primary flux calibrator (with assumed flux densities of 12.58 Jy at 2.1 GHz, 4.97 Jy at 5.5 GHz, 2.70 Jy at 9 GHz, and 1.11 Jy at 18 GHz) and PKSB0454$-$810 as the gain and phase calibrator. All calibrations were performed with the 2 GHz observing bands split into 8 bins.  After initial imaging, phase-only self-calibration was used to correct for atmospheric phase errors on timescales of a few minutes. We used multi-frequency synthesis in imaging and deconvolution and split the lower band into 3 sub-bands for imaging, centered at roughly 1.6 GHz, 2.1 GHz, and 2.7 GHz (the effective mean frequency of each sub-band varied slightly between epochs due to transient RFI).  At the lowest frequencies the entire primary beam was imaged to account for sidelobes of other sources in the field. Source fluxes were determined by fitting the point source response (gaussian clean beam) to the cleaned images. The later epochs exhibit clear fading relative to the initial observations (Figure \ref{fig:mcmc} top panels, Table \ref{tab:obs}). 

We investigated the consistency of the self-calibration across epochs by measuring the flux of a background object visible in each image, J073933.59$-$853954.3. There is no catalogued optical or radio source at this position, but a faint point-like source is detected in archival WISE observations obtained at a mean epoch of 2010 March 16. This object has a color of $W1-W2=-0.12\pm0.14$ mag, inconsistent with an AGN \citep{sab+12}, and shows no signs of infrared variability. We find that the radio flux of this second source changes by up to 10\% between epochs at all frequencies. These variations are 2-3 times larger than the image rms noise at 5.5 GHz and 9.0 GHz. Although it is possible that these changes are due to intrinsic variability of this source, we conservatively add an additional 10\% uncertainty to all flux densities in our modeling to account for possible calibration uncertainties. 

The location of XMMSL1 J0740$-$85 was also observed on 12 January 1998 and 24 October 1998 as part of the Sydney University Molonglo Sky Survey (SUMSS; \citealt{mmb+03}). No source was detected in a $10\arcmin$ by $10\arcmin$ combined image centered on the radio position down to a $5\sigma$ limit of 4.3 mJy at 843 MHz. If we assume no self-absorption and use a single power law to extrapolate our observed ATCA spectral energy distributions to 843 MHz, we find that even with this conservative assumption the source would have not been detected during any of our observations. The SUMSS limit therefore places only a very weak upper bound on the pre-flare radio variability of the source.

\subsection{Interstellar Scintillation}
Compact radio sources viewed through the interstellar medium (ISM) are observed to undergo random flux variations on timescales of hours to days. This effect, called interstellar scintillation, is caused by small-scale inhomogeneities in the ISM and can be significant at low radio frequencies. Using the NE2001 Galactic free electron energy density model \citep{cl02}, we find that the transition between strong and weak scintillation along our line of sight to XMMSL1 J0740$-$85 occurs at $\approx13$ GHz. Using the method of \cite{w98} and \cite{gn06}, we approximate the rms and typical timescale of the flux variations expected for a source of angular size 50 $\mu$as.\footnote{We choose 50 $\mu$as as a conservative estimate of the source size based on an initial fit to our epoch 1 observations that ignores any scintillation uncertainty; our subsequent analysis shows that including scintillation increases the uncertainty on our size estimate, but results in a similar value. See Section \ref{sec:nr}.} This size scale is comparable to the Fresnel scale at $\approx3$ GHz and the source can be treated as point-like below this frequency. In both the strong and the weak regimes, a point source will exhibit the strongest and most rapid flux variations. If the emitting region is larger than 50 $\mu$as, then scintillation effects will be further suppressed. 

From this model, we find that our 18 GHz observation is unlikely to be significantly affected by scintillation, with flux variations of $\lesssim 4\%$ and a timescale that is much shorter than our observation. Below 13 GHz, we expect both diffractive and refractive scintillation. Our observations are not sensitive to diffractive scintillation, which would require narrower bandwidths and shorter integration times to resolve \citep{w98}, but refractive scintillation is a broadband process and the timescales of the estimated flux variations are longer than our integration times. We estimate expected flux variations of $\sim15-40$\% between epochs, depending on the frequency (Table \ref{tab:obs}). This makes scintillation the dominant source of uncertainty in our measurements at low frequencies and we add the predicted scintillation variations in quadrature with the statistical and calibration uncertainties for all of our modeling. 

\section{Possible Origins of the Radio Emission}\label{sec:mod}
\subsection{Steady-State Processes}
We first consider whether the observed radio emission could be due to processes in the host galaxy unrelated to the TDE. The observed decline to $\sim60$\% of the original flux density over nine months is inconsistent with star formation. Furthermore, as discussed in \cite{sax16}, archival observations of the host galaxy reveal that it has little ongoing star formation activity and exhibited no signs of pre-TDE AGN activity. The host's optical spectrum showed no emission lines and archival GALEX observations restrict the current star formation rate to $\sim0.02$ M$_{\odot}$ yr$^{-1}$ \citep{sax16}. This star formation rate implies a radio flux density of $\sim 0.03$ mJy at 1.5 GHz \citep{con02}, which is a factor of 20 less than the flux density we observe in the last epoch. We therefore conclude that star formation contributes negligibly to the radio emission at all times probed by our observations. 

The flux decline rate is roughly consistent with the behavior of the radio AGN samples studied by \cite{hov08} and \cite{niep09}, who found that typical radio AGN flares took $\sim2$ years to decline back to quiescent flux levels. Each of our radio epochs can be fit by a single power law, $F_{\nu} \propto \nu^{-0.7 \pm 0.1}$. This spectral index is somewhat steeper than the typical flare spectra observed by \cite{hov08}, who found $F_{\nu} \propto \nu^{-0.24}$, but it is within the range of radio spectral indices observed in nearby Seyfert galaxies \citep{ho01}. The primary argument against an AGN origin for the radio emission thus comes from observations of the host at other wavelengths. Optical spectra of the host taken both before and after the TDE discovery showed none of the characteristic AGN emission lines and allowed \cite{sax16} to place an upper limit of $F_{\rm [OIII]}\lesssim 4\times10^{15}$ erg s$^{-1}$ cm$^{-2}$ on the flux of the [OIII]$\lambda$5007 line, which when combined with X-ray observations shows that the $L_{\text{2-10 keV}}/L_{\rm [OIII]}$ ratio of the galaxy is atypical for an AGN. The archival WISE galaxy colors are also consistent with a non-active galaxy \citep{sab+12,sax16}. We therefore conclude that all of the observed radio emission is associated with the TDE.

\subsection{Synchrotron Emission Model}

Our radio observations of XMMSL1 J0740$-$85 are broadly consistent with optically thin synchrotron emission. Below, we consider three possible scenarios for the origin of this emission in the context of a TDE. In all three scenarios, a blastwave generated by outflowing material accelerates the ambient electrons into a power law distribution $N(\gamma) \propto \gamma^{-p}$ for $\gamma \geq \gamma_m$, where $\gamma$ is the electron Lorentz factor, $\gamma_m$ is the minimum Lorentz factor of the distribution, and $p$ is the power law index. We follow the equipartition formalism outlined in \cite{dnp13}, which can be applied to both relativistic and non-relativistic outflows. This allows us to estimate the outflow energy ($E_{\rm eq}$) and the radius of the emitting region ($R_{\rm eq}$) by assuming that the the electron and magnetic field energy densities are near equipartition \citep{pac70,sr77,chev98}. We can then derive a number of other useful quantities, including the pre-existing circumnuclear density ($n$), the magnetic field strength ($B$), the outflow velocity ($v_{\rm ej}$, or $\beta_{\rm ej}$ when scaled to $c$), and the outflow mass ($M_{\rm ej}$). 

We note that this analysis relies on being able to identify a spectral peak ($\nu_p$), which corresponds to either the synchrotron frequency of electrons at $\gamma_m$ ($\nu_m$) or the self-absorption frequency ($\nu_a$), depending on the outflow parameters. For late-time observations like those considered here, we generically expect $\nu_m<\nu_a$ and therefore that $\nu_p=\nu_a$. This is true for both non-relativistic and initially relativistic outflows. If we assume $p=3$, as expected for a non-relativistic outflow \citep{dnp13}, we find that a Markov Chain Monte Carlo (MCMC) fitting technique can identify $\nu_p$ (Figure \ref{fig:mcmc}). This is possible because our data exhibit spectral flattening at low frequencies, allowing us to constrain the peak frequency even though the actual peak is near or just below the lower edge of our observing band. However, due to the additional uncertainty generated by scintillation, we cannot entirely rule out the possibility that $\nu_p$ is below our observing band for all three epochs (note the tail to low frequencies in all three epochs in the distributions shown in row 2 of Figure \ref{fig:mcmc}). If $p<3$, as expected for a relativistic outflow, the constraint on $\nu_p$ weakens further.

If the peak frequency has passed below the range of our observations, then we can still make progress by setting upper limits on $\nu_p$ and lower limits on the flux density of the peak ($F_{\nu,p}$). Since the outflow expands over time, we expect $\nu_p$ to evolve to lower frequencies, so the most constraining limit comes from the first epoch. The MCMC modeling gives $\nu_p=1.7\pm0.3$ GHz and $F_{\nu,p}=1.2\pm0.3$ mJy for this observation (Figure \ref{fig:mcmc}). For each of the models considered below, we therefore take $\nu_p\sim1.7$ GHz and $F_{\nu,p}\sim1.2$ mJy at a time $\Delta t\sim600$ days and make no attempt to discuss the time variation of these quantities.

\subsubsection{Relativistic Jet}

\begin{figure}
\centerline{\includegraphics[width=3.6in]{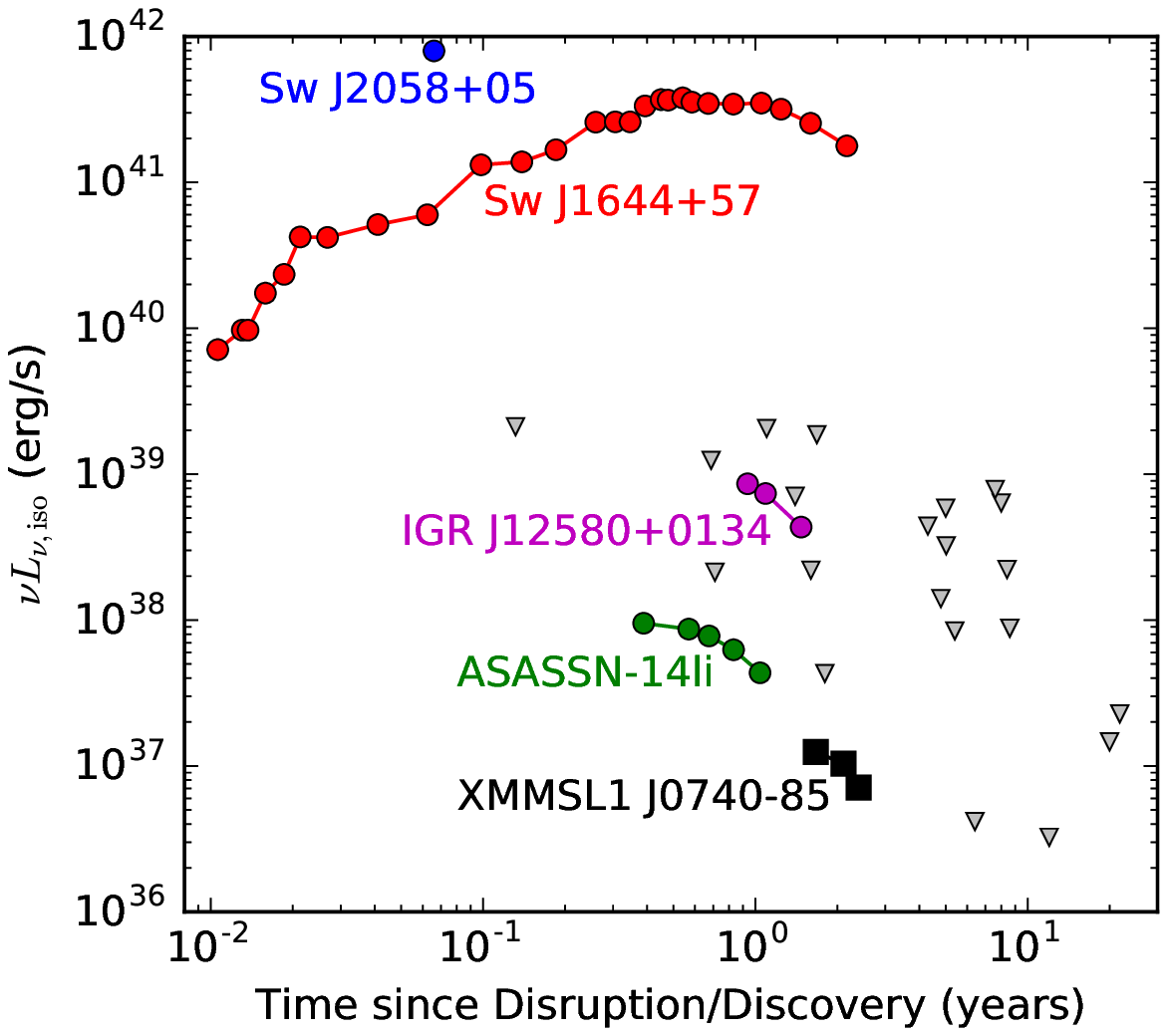}}
\caption{The radio luminosities of TDEs as a function of the time since disruption (or discovery date if a precise disruption time estimate is unavailable). Colored circles are literature detections for Sw J1644+57 \citep{zbs+11,bzp+12}, Sw J2058+05, \citep{ckh+12}, IGR J12580+0134 \citep{ir15}, and ASASSN-14li \citep{abg+16}. The luminosity of XMMSL1 J0740$-$85 is shown by the black squares. Gray triangles are $5\sigma$ upper limits \citep{kom02,bmc+13,vfk+13,cbg+14,ags+14}. The IGR J12580+0134 and ASASSN-14li points are the total radio luminosity observed during each flare and may include radio emission from processes unrelated to the TDE. All detected points are observations centered at 5-6 GHz, while the upper limits also include observations at 1.4 GHz, 3 GHz and 8.5 GHz.} \label{fig:lum}
\end{figure}

We first consider the possibility that the radio emission is caused by a relativistic jet launched during the phase of peak accretion onto the SMBH (assumed to coincide with the X-ray discovery date). The observed emission is orders of magnitude less luminous ($\nu L_{\nu} \sim10^{37}$ erg s$^{-1}$ at 5.5 GHz) than the on-axis relativistic jet seen in Sw J1644+57 at a similar time ($\nu L_{\nu} \sim10^{41}$ erg s$^{-1}$ at 5.8 GHz), so any jet in XMMSL1 J0740$-$85 must be much weaker (Figure \ref{fig:lum}). For any reasonable combination of parameters, an initially relativistic jet would have decelerated to non-relativistic velocities by the time of our first epoch \citep{np11}. The subsequent evolution of a decelerated jet is indistinguishable from that of a spherical, mildly-relativistic outflow, regardless of the initial orientation of the jet axis relative to our line of sight \citep{np11}. For all observing frequencies $\nu > \nu_m, \nu_a$, the light curve peaks at the deceleration time, $t_{\rm dec}\approx30E_{49}^{1/3}n^{-1/3}$ days, where $E_{49}$ is the jet energy in units of $10^{49}$ erg and $n$ is the density of the surrounding medium in units of cm$^{-3}$. At times $t>t_{\rm dec}$, the flux density at $\nu$ is given by $F_{\nu}(t) = F_{\nu,p}(t/t_{\rm dec})^{-(15p-21)/10}$, where $F_{\nu,p}\approx0.3E_{49} n^{(p+1)/4} \epsilon_{B,-1}^{(p+1)/4} \epsilon_{e,-1}^{p-1} d_{27}^{-2} (\nu/1.4 \text{ GHz})^{-(p-1)/2}$ mJy is the flux at $t_{\rm dec}$ \citep{np11}. Here, $\epsilon_e$ and $\epsilon_B$ are the fraction of the total energy carried by the electrons and by the magnetic field, respectively, and $d_{27}$ is the distance to the source in units of $10^{27}$ cm.

\begin{figure}
\centerline{\includegraphics[width=3.7in]{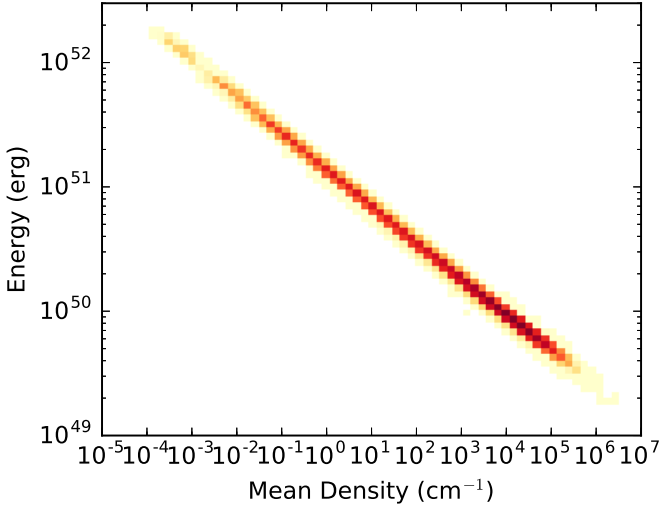}}
\caption{The distribution of circumnuclear densities and outflow energies allowed by our observations when assuming a decelerated relativistic jet model with a spectral index of $p=2.7$ for the accelerated electron population. This distribution was computed using the output of our MCMC modeling applied to our epoch 1 data. The degeneracy arises because $\nu_p$, the peak frequency of the radio spectral energy distribution, is only weakly constrained.} 
\label{fig:nvE}
\end{figure}

We observe a broadband flux decline throughout our observations, which implies that in this scenario $t_{\rm dec} \lesssim 600$ days and $F_{\nu,p} \gtrsim F_{\text{1.7 GHz}}(600$ days) $\sim 1.2$ mJy. By comparing the theoretical light curve $F_{\nu} \propto t^{-(15p-21)/10}$ to our observed light curve $F_{\nu} \propto t^{-2}$ between epochs 1 and 3 we find $p \sim2.7$. We assume that the system is in equipartition with $\epsilon_e=0.1$ and $\epsilon_B=6/11\epsilon_e$ \citep{dnp13}. This minimizes the total energy of the system. We can then use the above expressions for $t_{\rm dec}$, $F_{\nu,p}$, and $F_{\nu}(t)$ together with the output of our MCMC run with $p=2.7$ to determine the energy and circumnuclear density required to satisfy our observations. The resulting distribution of allowed energies and densities is shown in Figure \ref{fig:nvE}. We find that with 95\% confidence, the energy of the jet is between $5 \times 10^{49}$ erg and $4 \times 10^{51}$ erg and the density is between 0.03 cm$^{-3}$ and $7 \times 10^4$ cm$^{-3}$. The median density, $n=700$ cm$^{-3}$, is comparable to recent results that suggest typical densities in TDE host galaxies are $n \approx 0.5-2\times10^3$ cm$^{-3}$ at a distance of $10^{18}$ cm \citep{zbs+11,bzp+12,abg+16,gmm+16}. The median energy, $2 \times 10^{50}$ erg, is 100 times weaker than the $2 \times 10^{52}$ erg jet seen in Sw J1644+57 \citep{bzp+12}.

\subsubsection{Non-relativistic Outflow}\label{sec:nr}

We next model the radio emission as a non-relativistic outflow, using the same method applied to our radio observations of ASASSN-14li \citep{abg+16}. The primary model that we consider is a spherical outflow launched at the time of the X-ray discovery. This model is motivated by theoretical simulations that show a wind is expected during even mildly super-Eddington accretion, while jet formation may require more extreme conditions \citep{sq09,dc+12,tmg+14,ktn14}. We also consider a mildly collimated outflow with an angular cross-sectional area of $f_A=0.1$. We follow previous work \citep{dnp13,abg+16} and assume equipartition with $p=3$, $\epsilon_e=0.1$, and kinetic energy dominated by protons. We also assume that the emission peaks at the self-absorption frequency, synchrotron and Compton cooling are unimportant at our observing frequencies, and the emission emanates from a shell with a thickness of $0.1$ of the blastwave radius.

For $\nu_p\sim1.7$ GHz and $F_{\nu,p}\sim1.2$ mJy, we find that in the spherical case the outflow has a radius $R_{\rm eq}\sim5.1\times10^{16}$ cm and an energy $E_{\rm eq}\sim 1.5\times10^{48}$ erg. This implies an average expansion velocity of $v_{\rm ej} \sim10^4$ km s$^{-1}$ and an outflow mass of $M_{\rm ej} \sim 2\times10^{-3}$ M$_{\odot}$. We find that the average ambient density within $R_{\rm eq}$ is $n\sim100$ cm$^{-3}$, which means that the outflow has swept up an amount of material that is a negligible fraction of its total mass. We therefore expect that the outflow has not yet decelerated. Finally, we infer a moderate magnetic field strength $B\sim0.4$ G. This is an order of magnitude lower than the magnetic field strength inferred for Sw J1644+57 at early times \citep{zbs+11}. If the peak frequency of the radio spectral energy distribution is below our observing range in the first epoch, then the inferred values of $R_{\rm eq}$, $E_{\rm eq}$, $v_{\rm ej}$, and $M_{\rm ej}$ can be treated as lower limits while $n$ and $B$ can be treated as upper limits.

The mildly collimated outflow model gives similar results. The radius and velocity inferred are somewhat larger, $R_{\rm eq}\sim 1.5\times10^{17}$ cm and $v_{\rm ej}\sim 2.9 \times 10^4$ km s$^{-1}$, but this is still consistent with a non-relativistic treatment. The energy and mass of the outflow are somewhat lower, $E_{\rm eq}\sim 6\times10^{47}$ erg and $M_{\rm ej}\sim8\times10^{-5}$ M$_{\odot}$, as are the average ambient density, $n\sim60$ cm$^{-3}$, and the magnetic field strength, $B\sim0.2$ G. For both models, these properties are similar to those of the non-relativistic outflow found in ASSASN-14li \citep{abg+16}, which would make XMMSL1 J0740$-$85 the second known TDE with this less energetic type of outflow.

\begin{figure}
\centerline{\includegraphics[width=3.7in]{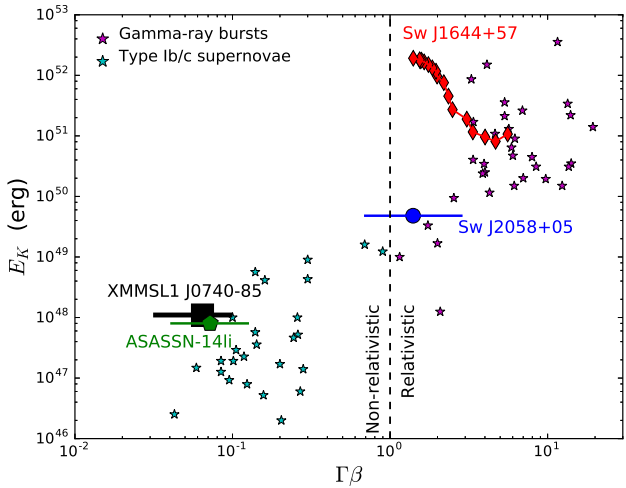}}
\caption{Kinetic energy ($E_K$) as a function of outflow velocity ($\Gamma\beta$) from radio observations of TDEs. We show the inferred values for our non-relativistic XMMSL1 J0740$-$85 model (black square; horizontal bar represents the range of velocity for a range of outflow geometries) as well as ASASSN-14li (green pentagon; \citealt{abg+16}) and the two $\gamma$-ray TDEs with radio emission: Sw\,J1644+57 (red diamonds; \citealt{zbs+11} and \citealt{bzp+12}) and Sw\,J2058+05 (blue circle; \citealt{ckh+12}).  The data for Sw\,J1644+57 are from detailed modeling of the radio emission as a function of time, including a correction for jet collimation with an opening angle of about 0.1 rad \citep{zbs+11,bzp+12}.  The data points and velocity ranges for Sw\,J2058+05 and ASASSN-14li are based on an identical analysis to the one carried out here \citep{abg+16}. Also shown for comparison are a sample of long-duration $\gamma$-ray bursts (LGRBs; magenta stars) and Type Ib/c core-collapse supernovae (Type Ib/c SNe; cyan stars) \citep{mms+14}. } 
\label{fig:ek}
\end{figure}

\subsubsection{Unbound Debris}
When a star is tidally disrupted, approximately half of the debris will ultimately accrete onto the black hole, while the rest is unbound \citep{rees88}. We consider whether the observed emission could be due to the interaction between the unbound debris and the circumnuclear medium \citep{km96}. We expect the velocity of the unbound debris to be $\sim 10^4$ km s$^{-1}$, so a non-relativistic model similar to that considered in the previous section is appropriate. However, the size of the emitting region will be much smaller, as simulations have shown that the unbound debris stream is expected to be initially self-gravitating for all but the most extreme event geometries \citep{k94,gmr14,cn15}. In this case, the solid angle subtended by the unbound debris will decrease as the stream leaves the vicinity of the SMBH and will only begin homologous expansion at a distance of $\sim10^{16}$ cm. At this distance, the stream will cover a solid angle of $\sim10^{-5}$ steradians \citep{gmc15} and any radio emission produced will be orders of magnitude too faint to explain our observed radio emission. 

For non self-gravitating streams, (created by events in which the disrupted star's closest point of approach to the SMBH is $\lesssim1/3$ of the tidal radius), the solid angle subtended by the stream is determined by the spread in velocity of the unbound debris and is roughly 0.2 steradians for a non-spinning $10^6$ M$_{\odot}$ black hole \citep{sq09}. For spinning black holes, the velocity spread may increase or decrease by up to a factor of 2 \citep{k12}. Streams with a low radiative efficiency may also be non self-gravitating; \cite{krolik16} suggested that a bow shock between the unbound debris and the ambient medium could heat the stream beyond its ability to cool, increasing the size of the emitting region enough to explain the radio emission of ASASSN-14li. However, this model requires a high circumnuclear density and is sensitive to the velocity distribution of the unbound debris. Repeating our non-relativistic analysis from the previous section for a solid angle of 0.2 steradians ($f_A=0.063$), we find that XMMSL1 J0740$-$85's radio emission can be explained by outflowing material at a radius of $R_{\rm eq}\sim1.9\times10^{17}$ with an average velocity $v_{\rm ej} \sim 3.6\times10^4$ km s$^{-1}$ interacting with a circumnuclear medium with an average density of $n\sim50$ cm$^{-3}$. Since the inferred mass is small, $M_{\rm ej}\sim 5\times 10^{-5}$ M$_\odot$, this means that we are not observing radio emission from the entire unbound debris stream. This could be plausible if we are only seeing the fastest-moving material at the leading edge of the unbound debris stream (as suggested for ASASSN-14li by \citealt{krolik16}), but due to the rarity of such close star-SMBH encounters, we consider emission from a non self-gravitating unbound debris stream to be a less likely explanation for the radio emission. 

\section{Discussion}\label{sec:disc}

Our observations make XMMSL1 J0740$-$85 the fifth TDE with detected radio emission. Even with such a small sample size, it is clear that there is a wide diversity in the radio properties of TDEs (Figure \ref{fig:lum}). The clearest distinction is between TDEs that produce luminous relativistic jets, like Sw J1644+57 and Sw J2058+05, and TDEs that produce much weaker emission, like ASASSN-14li (Figure \ref{fig:ek}). The recent TDE candidate IGR J12580+0134 has a radio luminosity between these two extremes, but could have launched an off-axis jet as powerful as Sw J1644+57 \citep{ir15,lei16}. Complicating the analysis of this event, the host of IGR J12580+0134 is a known AGN and had a pre-flare radio luminosity $\sim6$ times fainter than the peak luminosity reached during the flare \citep{ir15}. Even the most variable AGN rarely undergo flux changes of this magnitude at 6 GHz, but further study may be needed to disentangle the TDE from other AGN activity. We attempt no further analysis of IGR J12580+0134 in this paper.

Our radio observations of XMMSL1 J0740$-$85 are unable to directly distinguish between a decelerated weak relativistic jet and a non-relativistic outflow model, but they do require any jet in XMMSL1 J0740$-$85 to be much less energetic than the jet seen in Sw J1644+57. The similar energy scales inferred from the radio observations imply that XMMSL1 J0740$-$85 has more in common with ASASSN-14li than with the relativistic events, which may suggest that the non-relativistic outflow model considered here is more appropriate than a jet. Furthermore, while the relativistic {\it Swift} events were highly super-Eddington, the peak accretion rate inferred from X-ray observations of XMMSL1 J0740$-$85 is mildly sub-Eddington \citep{sax16}. This is also similar to ASASSN-14li, where modeling of the X-ray, UV, and optical emission showed that this event was at most only mildly super-Eddington \citep{mkm+15,hkp+16,abg+16}. 

Extreme jetted TDEs exhibit $\gamma$-ray emission and relativistic outflows with a large kinetic energy, but they represent at most a few percent of the overall TDE volumetric rate \citep{mgm+15}. On the other hand, events like XMMSL1 J0740$-$85 and ASASSN-14li exhibit less energetic outflows and appear to represent the bulk of the TDE population \citep{abg+16}. Published upper limits on radio emission from 15 archival events can rule out Sw J1644+57-like jets in many cases \citep{kom02,bmc+13,vfk+13,cbg+14,ags+14}, but the discovery of XMMSL1 J0740$-$85 reinforces the idea that many of the more distant literature TDEs could have also produced radio emission at a luminosity too low to be detectable with current facilities (Figure \ref{fig:lum}). The TDE sample, although small, appears to trace the same relation seen in LGRBs and Type Ib/c SNe (Figure \ref{fig:ek}). The LGRBs exhibit relativistic outflows with $E_K\simgt 10^{50}$ erg, while Type Ib/c SNe have non-relativistic outflows with $E_K\simlt 10^{49}$ erg. In addition, LGRBs represent $\simlt 1\%$ of the Type Ib/c SN rate \citep{wp10}. 

\begin{figure}
\centerline{\includegraphics[width=3.55in]{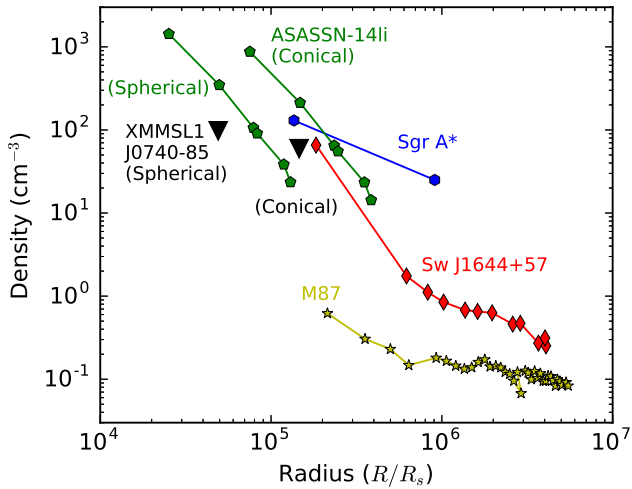}}
\caption{The average density in the circumnuclear region of XMMSL1 J0740$-$85 (black triangles), as computed for our two non-relativistic outflow models of the radio emission (a spherical outflow and a mildly collimated outflow with $f_A=0.1$). If the radio flux peak is below our observing frequencies, these points become upper limits. For comparison, we show the density profiles for Sgr A$^*$ \citep{bmm+03}, M87, \citep{rfm+15}, the $\gamma$-ray TDE Sw\,J1644+57 \citep{bzp+12}, and the non-relativistic TDE ASASSN-14li \citep{abg+16}. To facilitate the comparison we scale the radii by the Schwarzschild radius of each SMBH ($R_s$), taking $M_{\rm BH}\approx 3.5\times10^6$ M$_\odot$ for XMMSL1 J0740$-$85 \citep{sax16}.  We find that  the density of the circumnuclear region of XMMSL1 J0740$-$85 is comparable to the other SMBH systems.}
\label{fig:n}
\end{figure} 

Radio observations of TDEs are also rapidly becoming a vital tool to study the population of quiescent SMBHs in nearby galaxies, as they probe the density around SMBHs at otherwise unresolvable parsec and sub-parsec scales. Comparable resolution has been recently achieved for ASASSN-14li using infrared observations of the dust emission from the host nucleus, which reveal a light echo from the flare \citep{lke16,vmk+16}, but otherwise is only directly measurable for the SMBH in our own galaxy, Sagittarius A* \citep{bmm+03}, and for the $\sim5\times10^9$ M$_{\odot}$ SMBH in M87 if we scale by the black hole's Schwarzschild radius ($R_s=2GM_{\rm BH}/c^2$, where $M_{\rm BH}$ is the black hole mass). We show the density inferred from our non-relativistic outflow model of XMMSL1 J0740$-$85 in comparison with the circumnuclear density profiles derived from other TDE radio observations in Figure \ref{fig:n}. We see that for a range of plausible outflow geometries, the density at the core of XMMSL1 J0740$-$85's host galaxy is comparable to that seen around ASASSN-14li, Sw J1644+57, and Sgr A* when scaled by the Schwarzschild radius (and therefore by the mass) of each SMBH.

\section{Conclusions}\label{sec:conc}

We have analyzed radio emission localized to the nucleus of the host galaxy of the TDE candidate XMMSL1 J0740$-$85 \citep{sax16}. We find that the radio emission is consistent with a non-relativistic outflow that has similar properties to the outflow discovered in ASASSN-14li \citep{abg+16}, making XMMSL1 J0740$-$85 only the second TDE known to produce radio emission of this type. Other explanations such as a weak initially-relativistic jet or emission from the unbound debris generated by a deeply penetrating tidal encounter are also possible, but less likely. A strong relativistic jet like that seen in Sw J1644+57 is ruled out. Our radio observations of XMMSL1 J0740$-$85 point to the importance of TDE radio studies, but also highlight the importance of early observations to constrain the overall energy scale while the ambient density is still high enough for the self-absorption peak to be visible in the radio band.

With an ever-increasing number of optical, X-ray, and radio surveys slated to discover tens to hundreds of new TDEs per year over the coming decades, we expect to discover radio emission from many more jetted and non-jetted TDEs. An event with the radio luminosity of XMMSL1 J0740$-$85 ($L_{\nu} \sim 3\times 10^{27}$ erg s$^{-1}$ Hz$^{-1}$ at 5.5 GHz) can already be detected out to a distance of $\sim230$ Mpc with a single ATCA observation and $\sim 300$ Mpc with a one-hour VLA observation. Our observations of XMMSL1 J0740$-$85 are an important step towards more fully characterizing outflows in TDEs and the detailed properties of the circumnuclear environments of SMBHs.

\begin{acknowledgements} We thank the anonymous referee for helpful comments that have improved this manuscript. K.D.A.~and E.B.~acknowledge partial support from the NSF under grant AST-1411763 and from NASA under grant NNX15AE50G. We thank Phil Edwards for rapidly scheduling our first epoch of ATCA observations. The Australia Telescope Compact Array is part of the Australia Telescope National Facility which is funded by the Australian Government for operation as a National Facility managed by CSIRO. \end{acknowledgements}

\software{Astropy \citep{astro13}, Matplotlib \citep{hunter07}, Miriad \citep{stw95}, NumPy \citep{vcv11}, SciPy \citep{scipy}}

\bibliographystyle{aasjournal}
\bibliography{radio}

\begin{thebibliography}{}
\expandafter\ifx\csname natexlab\endcsname\relax\def\natexlab#1{#1}\fi

\bibitem[{{Alexander} {et~al.}(2016){Alexander}, {Berger}, {Guillochon},
  {Zauderer}, \& {Williams}}]{abg+16}
{Alexander}, K.~D., {Berger}, E., {Guillochon}, J., {Zauderer}, B.~A., \&
  {Williams}, P.~K.~G. 2016, \apjl, 819, L25

\bibitem[{{Arcavi} {et~al.}(2014){Arcavi}, {Gal-Yam}, {Sullivan}, {Pan},
  {Cenko}, {Horesh}, {Ofek}, {De Cia}, {Yan}, {Yang}, {Howell}, {Tal},
  {Kulkarni}, {Tendulkar}, {Tang}, {Xu}, {Sternberg}, {Cohen}, {Bloom},
  {Nugent}, {Kasliwal}, {Perley}, {Quimby}, {Miller}, {Theissen}, \&
  {Laher}}]{ags+14}
{Arcavi}, I., {Gal-Yam}, A., {Sullivan}, M., {et~al.} 2014, \apj, 793, 38

\bibitem[{{Astropy Collaboration} {et~al.}(2013){Astropy Collaboration},
  {Robitaille}, {Tollerud}, {Greenfield}, {Droettboom}, {Bray}, {Aldcroft},
  {Davis}, {Ginsburg}, {Price-Whelan}, {Kerzendorf}, {Conley}, {Crighton},
  {Barbary}, {Muna}, {Ferguson}, {Grollier}, {Parikh}, {Nair}, {Unther},
  {Deil}, {Woillez}, {Conseil}, {Kramer}, {Turner}, {Singer}, {Fox}, {Weaver},
  {Zabalza}, {Edwards}, {Azalee Bostroem}, {Burke}, {Casey}, {Crawford},
  {Dencheva}, {Ely}, {Jenness}, {Labrie}, {Lim}, {Pierfederici}, {Pontzen},
  {Ptak}, {Refsdal}, {Servillat}, \& {Streicher}}]{astro13}
{Astropy Collaboration}, {Robitaille}, T.~P., {Tollerud}, E.~J., {et~al.} 2013,
  \aap, 558, A33

\bibitem[{{Baganoff} {et~al.}(2003){Baganoff}, {Maeda}, {Morris}, {Bautz},
  {Brandt}, {Cui}, {Doty}, {Feigelson}, {Garmire}, {Pravdo}, {Ricker}, \&
  {Townsley}}]{bmm+03}
{Baganoff}, F.~K., {Maeda}, Y., {Morris}, M., {et~al.} 2003, ApJ, 591, 891

\bibitem[{{Barniol Duran} {et~al.}(2013){Barniol Duran}, {Nakar}, \&
  {Piran}}]{dnp13}
{Barniol Duran}, R., {Nakar}, E., \& {Piran}, T. 2013, ApJ, 772, 78

\bibitem[{{Berger} {et~al.}(2012){Berger}, {Zauderer}, {Pooley}, {Soderberg},
  {Sari}, {Brunthaler}, \& {Bietenholz}}]{bzp+12}
{Berger}, E., {Zauderer}, A., {Pooley}, G.~G., {et~al.} 2012, ApJ, 748, 36

\bibitem[{{Bloom} {et~al.}(2011){Bloom}, {Giannios}, {Metzger}, {Cenko},
  {Perley}, {Butler}, {Tanvir}, {Levan}, {O'Brien}, {Strubbe}, {De Colle},
  {Ramirez-Ruiz}, {Lee}, {Nayakshin}, {Quataert}, {King}, {Cucchiara},
  {Guillochon}, {Bower}, {Fruchter}, {Morgan}, \& {van der Horst}}]{bgm+11}
{Bloom}, J.~S., {Giannios}, D., {Metzger}, B.~D., {et~al.} 2011, Science, 333,
  203

\bibitem[{{Bower} {et~al.}(2013){Bower}, {Metzger}, {Cenko}, {Silverman}, \&
  {Bloom}}]{bmc+13}
{Bower}, G.~C., {Metzger}, B.~D., {Cenko}, S.~B., {Silverman}, J.~M., \&
  {Bloom}, J.~S. 2013, ApJ, 763, 84

\bibitem[{{Brown} {et~al.}(2015){Brown}, {Levan}, {Stanway}, {Tanvir}, {Cenko},
  {Berger}, {Chornock}, \& {Cucchiaria}}]{bls+15}
{Brown}, G.~C., {Levan}, A.~J., {Stanway}, E.~R., {et~al.} 2015, \mnras, 452,
  4297

\bibitem[{{Burrows} {et~al.}(2011){Burrows}, {Kennea}, {Ghisellini}, {Mangano},
  {Zhang}, {Page}, {Eracleous}, {Romano}, {Sakamoto}, {Falcone}, {Osborne},
  {Campana}, {Beardmore}, {Breeveld}, {Chester}, {Corbet}, {Covino},
  {Cummings}, {D'Avanzo}, {D'Elia}, {Esposito}, {Evans}, {Fugazza}, {Gelbord},
  {Hiroi}, {Holland}, {Huang}, {Im}, {Israel}, {Jeon}, {Jeon}, {Jun}, {Kawai},
  {Kim}, {Krimm}, {Marshall}, {P.~M{\'e}sz{\'a}ros}, {Negoro}, {Omodei},
  {Park}, {Perkins}, {Sugizaki}, {Sung}, {Tagliaferri}, {Troja}, {Ueda},
  {Urata}, {Usui}, {Antonelli}, {Barthelmy}, {Cusumano}, {Giommi}, {Melandri},
  {Perri}, {Racusin}, {Sbarufatti}, {Siegel}, \& {Gehrels}}]{bkg+11}
{Burrows}, D.~N., {Kennea}, J.~A., {Ghisellini}, G., {et~al.} 2011, Nature,
  476, 421

\bibitem[{{Cenko} {et~al.}(2012){Cenko}, {Krimm}, {Horesh}, {Rau}, {Frail},
  {Kennea}, {Levan}, {Holland}, {Butler}, {Quimby}, {Bloom}, {Filippenko},
  {Gal-Yam}, {Greiner}, {Kulkarni}, {Ofek}, {Olivares E.}, {Schady},
  {Silverman}, {Tanvir}, \& {Xu}}]{ckh+12}
{Cenko}, S.~B., {Krimm}, H.~A., {Horesh}, A., {et~al.} 2012, ApJ, 753, 77

\bibitem[{{Chevalier}(1998)}]{chev98}
{Chevalier}, R.~A. 1998, ApJ, 499, 810

\bibitem[{{Chornock} {et~al.}(2014){Chornock}, {Berger}, {Gezari}, {Zauderer},
  {Rest}, {Chomiuk}, {Kamble}, {Soderberg}, {Czekala}, {Dittmann}, {Drout},
  {Foley}, {Fong}, {Huber}, {Kirshner}, {Lawrence}, {Lunnan}, {Marion},
  {Narayan}, {Riess}, {Roth}, {Sanders}, {Scolnic}, {Smartt}, {Smith},
  {Stubbs}, {Tonry}, {Burgett}, {Chambers}, {Flewelling}, {Hodapp}, {Kaiser},
  {Magnier}, {Martin}, {Neill}, {Price}, \& {Wainscoat}}]{cbg+14}
{Chornock}, R., {Berger}, E., {Gezari}, S., {et~al.} 2014, ApJ, 780, 44

\bibitem[{{Condon} {et~al.}(2002){Condon}, {Cotton}, \& {Broderick}}]{con02}
{Condon}, J.~J., {Cotton}, W.~D., \& {Broderick}, J.~J. 2002, \aj, 124, 675

\bibitem[{{Cordes} \& {Lazio}(2002)}]{cl02}
{Cordes}, J.~M., \& {Lazio}, T.~J.~W. 2002, ArXiv Astrophysics e-prints,
  astro-ph/0207156

\bibitem[{{Coughlin} \& {Nixon}(2015)}]{cn15}
{Coughlin}, E.~R., \& {Nixon}, C. 2015, \apjl, 808, L11

\bibitem[{{De Colle} {et~al.}(2012){De Colle}, {Guillochon}, {Naiman}, \&
  {Ramirez-Ruiz}}]{dc+12}
{De Colle}, F., {Guillochon}, J., {Naiman}, J., \& {Ramirez-Ruiz}, E. 2012,
  \apj, 760, 103

\bibitem[{{French} {et~al.}(2016){French}, {Arcavi}, \& {Zabludoff}}]{faz16}
{French}, K.~D., {Arcavi}, I., \& {Zabludoff}, A. 2016, \apjl, 818, L21

\bibitem[{{Generozov} {et~al.}(2016){Generozov}, {Mimica}, {Metzger}, {Stone},
  {Giannios}, \& {Aloy}}]{gmm+16}
{Generozov}, A., {Mimica}, P., {Metzger}, B.~D., {et~al.} 2016, ArXiv e-prints,
  arXiv:1605.08437

\bibitem[{{Giannios} \& {Metzger}(2011)}]{gm11}
{Giannios}, D., \& {Metzger}, B.~D. 2011, \mnras, 416, 2102

\bibitem[{{Goodman} \& {Narayan}(2006)}]{gn06}
{Goodman}, J., \& {Narayan}, R. 2006, ApJ, 636, 510

\bibitem[{{Guillochon} {et~al.}(2014){Guillochon}, {Manukian}, \&
  {Ramirez-Ruiz}}]{gmr14}
{Guillochon}, J., {Manukian}, H., \& {Ramirez-Ruiz}, E. 2014, ApJ, 783, 23

\bibitem[{{Guillochon} {et~al.}(2015){Guillochon}, {McCourt}, {Chen},
  {Johnson}, \& {Berger}}]{gmc15}
{Guillochon}, J., {McCourt}, M., {Chen}, X., {Johnson}, M.~D., \& {Berger}, E.
  2015, ArXiv e-prints, arXiv:1509.08916

\bibitem[{{Guillochon} \& {Ramirez-Ruiz}(2013)}]{gr13}
{Guillochon}, J., \& {Ramirez-Ruiz}, E. 2013, ApJ, 767, 25

\bibitem[{Ho \& Ulvestad(2001)}]{ho01}
Ho, L.~C., \& Ulvestad, J.~S. 2001, The Astrophysical Journal Supplement
  Series, 133, 77

\bibitem[{{Holoien} {et~al.}(2016){Holoien}, {Kochanek}, {Prieto}, {Stanek},
  {Dong}, {Shappee}, {Grupe}, {Brown}, {Basu}, {Beacom}, {Bersier},
  {Brimacombe}, {Danilet}, {Falco}, {Guo}, {Jose}, {Herczeg}, {Long},
  {Pojmanski}, {Simonian}, {Szczygie{\l}}, {Thompson}, {Thorstensen}, {Wagner},
  \& {Wo{\'z}niak}}]{hkp+16}
{Holoien}, T.~W.-S., {Kochanek}, C.~S., {Prieto}, J.~L., {et~al.} 2016, \mnras,
  455, 2918

\bibitem[{{Hovatta} {et~al.}(2008){Hovatta}, {Nieppola}, {Tornikoski},
  {Valtaoja}, {Aller}, \& {Aller}}]{hov08}
{Hovatta}, T., {Nieppola}, E., {Tornikoski}, M., {et~al.} 2008, A\&A, 485, 51

\bibitem[{Hunter(2007)}]{hunter07}
Hunter, J.~D. 2007, Computing In Science \& Engineering, 9, 90

\bibitem[{{Irwin} {et~al.}(2015){Irwin}, {Henriksen}, {Krause}, {Wang},
  {Wiegert}, {Murphy}, {Heald}, \& {Perlman}}]{ir15}
{Irwin}, J.~A., {Henriksen}, R.~N., {Krause}, M., {et~al.} 2015, \apj, 809, 172

\bibitem[{Jones {et~al.}(2001--)Jones, Oliphant, Peterson, {et~al.}}]{scipy}
Jones, E., Oliphant, T., Peterson, P., {et~al.} 2001--, {SciPy}: Open source
  scientific tools for {Python}, , , [Online; accessed 2017-02-07]

\bibitem[{{Kelley} {et~al.}(2014){Kelley}, {Tchekhovskoy}, \&
  {Narayan}}]{ktn14}
{Kelley}, L.~Z., {Tchekhovskoy}, A., \& {Narayan}, R. 2014, \mnras, 445, 3919

\bibitem[{{Kesden}(2012)}]{k12}
{Kesden}, M. 2012, \prd, 86, 064026

\bibitem[{{Khokhlov} \& {Melia}(1996)}]{km96}
{Khokhlov}, A., \& {Melia}, F. 1996, \apjl, 457, L61

\bibitem[{{Kochanek}(1994)}]{k94}
{Kochanek}, C.~S. 1994, \apj, 422, 508

\bibitem[{{Komossa}(2002)}]{kom02}
{Komossa}, S. 2002, in Reviews in Modern Astronomy, Vol.~15, Reviews in Modern
  Astronomy, ed. R.~E. {Schielicke}, 27

\bibitem[{{Komossa}(2015)}]{kom15}
{Komossa}, S. 2015, Journal of High Energy Astrophysics, 7, 148

\bibitem[{{Krolik} {et~al.}(2016){Krolik}, {Piran}, {Svirski}, \&
  {Cheng}}]{krolik16}
{Krolik}, J., {Piran}, T., {Svirski}, G., \& {Cheng}, R.~M. 2016, \apj, 827,
  127

\bibitem[{{Lei} {et~al.}(2016){Lei}, {Yuan}, {Zhang}, \& {Wang}}]{lei16}
{Lei}, W.-H., {Yuan}, Q., {Zhang}, B., \& {Wang}, D. 2016, \apj, 816, 20

\bibitem[{{Levan} {et~al.}(2011){Levan}, {Tanvir}, {Cenko}, {Perley},
  {Wiersema}, {Bloom}, {Fruchter}, {Postigo}, {O'Brien}, {Butler}, {van der
  Horst}, {Leloudas}, {Morgan}, {Misra}, {Bower}, {Farihi}, {Tunnicliffe},
  {Modjaz}, {Silverman}, {Hjorth}, {Th{\"o}ne}, {Cucchiara}, {Cer{\'o}n},
  {Castro-Tirado}, {Arnold}, {Bremer}, {Brodie}, {Carroll}, {Cooper}, {Curran},
  {Cutri}, {Ehle}, {Forbes}, {Fynbo}, {Gorosabel}, {Graham}, {Hoffman},
  {Guziy}, {Jakobsson}, {Kamble}, {Kerr}, {Kasliwal}, {Kouveliotou},
  {Kocevski}, {Law}, {Nugent}, {Ofek}, {Poznanski}, {Quimby}, {Rol},
  {Romanowsky}, {S{\'a}nchez-Ram{\'{\i}}rez}, {Schulze}, {Singh}, {van
  Spaandonk}, {Starling}, {Strom}, {Tello}, {Vaduvescu}, {Wheatley}, {Wijers},
  {Winters}, \& {Xu}}]{ltc+11}
{Levan}, A.~J., {Tanvir}, N.~R., {Cenko}, S.~B., {et~al.} 2011, Science, 333,
  199

\bibitem[{{Lu} {et~al.}(2016){Lu}, {Kumar}, \& {Evans}}]{lke16}
{Lu}, W., {Kumar}, P., \& {Evans}, N.~J. 2016, \mnras, 458, 575

\bibitem[{{Margutti} {et~al.}(2014){Margutti}, {Milisavljevic}, {Soderberg},
  {Guidorzi}, {Morsony}, {Sanders}, {Chakraborti}, {Ray}, {Kamble}, {Drout},
  {Parrent}, {Zauderer}, \& {Chomiuk}}]{mms+14}
{Margutti}, R., {Milisavljevic}, D., {Soderberg}, A.~M., {et~al.} 2014, ApJ,
  797, 107

\bibitem[{{Mauch} {et~al.}(2003){Mauch}, {Murphy}, {Buttery}, {Curran},
  {Hunstead}, {Piestrzynski}, {Robertson}, \& {Sadler}}]{mmb+03}
{Mauch}, T., {Murphy}, T., {Buttery}, H.~J., {et~al.} 2003, \mnras, 342, 1117

\bibitem[{{Miller} {et~al.}(2015){Miller}, {Kaastra}, {Miller}, {Reynolds},
  {Brown}, {Cenko}, {Drake}, {Gezari}, {Guillochon}, {Gultekin}, {Irwin},
  {Levan}, {Maitra}, {Maksym}, {Mushotzky}, {O'Brien}, {Paerels}, {de Plaa},
  {Ramirez-Ruiz}, {Strohmayer}, \& {Tanvir}}]{mkm+15}
{Miller}, J.~M., {Kaastra}, J.~S., {Miller}, M.~C., {et~al.} 2015, \nat, 526,
  542

\bibitem[{{Mimica} {et~al.}(2015){Mimica}, {Giannios}, {Metzger}, \&
  {Aloy}}]{mgm+15}
{Mimica}, P., {Giannios}, D., {Metzger}, B.~D., \& {Aloy}, M.~A. 2015, MNRAS,
  450, 2824

\bibitem[{{Nakar} \& {Piran}(2011)}]{np11}
{Nakar}, E., \& {Piran}, T. 2011, \nat, 478, 82

\bibitem[{Nieppola {et~al.}(2009)Nieppola, Hovatta, Tornikoski, Valtaoja,
  Aller, \& Aller}]{niep09}
Nieppola, E., Hovatta, T., Tornikoski, M., {et~al.} 2009, The Astronomical
  Journal, 137, 5022

\bibitem[{{Pacholczyk}(1970)}]{pac70}
{Pacholczyk}, A.~G. 1970, {Radio astrophysics. Nonthermal processes in galactic
  and extragalactic sources}

\bibitem[{{Rees}(1988)}]{rees88}
{Rees}, M.~J. 1988, Nature, 333, 523

\bibitem[{{Russell} {et~al.}(2015){Russell}, {Fabian}, {McNamara}, \&
  {Broderick}}]{rfm+15}
{Russell}, H.~R., {Fabian}, A.~C., {McNamara}, B.~R., \& {Broderick}, A.~E.
  2015, MNRAS, 451, 588

\bibitem[{{Sault} {et~al.}(1995){Sault}, {Teuben}, \& {Wright}}]{stw95}
{Sault}, R.~J., {Teuben}, P.~J., \& {Wright}, M.~C.~H. 1995, in Astronomical
  Society of the Pacific Conference Series, Vol.~77, Astronomical Data Analysis
  Software and Systems IV, ed. R.~A. {Shaw}, H.~E. {Payne}, \& J.~J.~E.
  {Hayes}, 433

\bibitem[{{Saxton} {et~al.}(2008){Saxton}, {Read}, {Esquej}, {Freyberg},
  {Altieri}, \& {Bermejo}}]{sax08}
{Saxton}, R.~D., {Read}, A.~M., {Esquej}, P., {et~al.} 2008, \aap, 480, 611

\bibitem[{{Saxton} {et~al.}(2016){Saxton}, {Read}, {Komossa}, {Lira},
  {Alexander}, \& {Wieringa}}]{sax16}
{Saxton}, R.~D., {Read}, A.~M., {Komossa}, S., {et~al.} 2016, \aap,
  arXiv:1610.01788

\bibitem[{{Scott} \& {Readhead}(1977)}]{sr77}
{Scott}, M.~A., \& {Readhead}, A.~C.~S. 1977, MNRAS, 180, 539

\bibitem[{{Stern} {et~al.}(2012){Stern}, {Assef}, {Benford}, {Blain}, {Cutri},
  {Dey}, {Eisenhardt}, {Griffith}, {Jarrett}, {Lake}, {Masci}, {Petty},
  {Stanford}, {Tsai}, {Wright}, {Yan}, {Harrison}, \& {Madsen}}]{sab+12}
{Stern}, D., {Assef}, R.~J., {Benford}, D.~J., {et~al.} 2012, \apj, 753, 30

\bibitem[{{Strubbe} \& {Quataert}(2009)}]{sq09}
{Strubbe}, L.~E., \& {Quataert}, E. 2009, MNRAS, 400, 2070

\bibitem[{{Tchekhovskoy} {et~al.}(2014){Tchekhovskoy}, {Metzger}, {Giannios},
  \& {Kelley}}]{tmg+14}
{Tchekhovskoy}, A., {Metzger}, B.~D., {Giannios}, D., \& {Kelley}, L.~Z. 2014,
  \mnras, 437, 2744

\bibitem[{van~der Walt {et~al.}(2011)van~der Walt, Colbert, \&
  Varoquaux}]{vcv11}
van~der Walt, S., Colbert, S.~C., \& Varoquaux, G. 2011, Computing in Science
  \& Engineering, 13, 22

\bibitem[{{van Velzen} {et~al.}(2013){van Velzen}, {Frail}, {K{\"o}rding}, \&
  {Falcke}}]{vfk+13}
{van Velzen}, S., {Frail}, D.~A., {K{\"o}rding}, E., \& {Falcke}, H. 2013,
  A\&A, 552, A5

\bibitem[{{van Velzen} {et~al.}(2011){van Velzen}, {K{\"o}rding}, \&
  {Falcke}}]{vkf11}
{van Velzen}, S., {K{\"o}rding}, E., \& {Falcke}, H. 2011, \mnras, 417, L51

\bibitem[{{van Velzen} {et~al.}(2016{\natexlab{a}}){van Velzen}, {Mendez},
  {Krolik}, \& {Gorjian}}]{vmk+16}
{van Velzen}, S., {Mendez}, A.~J., {Krolik}, J.~H., \& {Gorjian}, V.
  2016{\natexlab{a}}, \apj, 829, 19

\bibitem[{{van Velzen} {et~al.}(2016{\natexlab{b}}){van Velzen}, {Anderson},
  {Stone}, {Fraser}, {Wevers}, {Metzger}, {Jonker}, {van der Horst}, {Staley},
  {Mendez}, {Miller-Jones}, {Hodgkin}, {Campbell}, \& {Fender}}]{vas+16}
{van Velzen}, S., {Anderson}, G.~E., {Stone}, N.~C., {et~al.}
  2016{\natexlab{b}}, Science, 351, 62

\bibitem[{{Walker}(1998)}]{w98}
{Walker}, M.~A. 1998, MNRAS, 294, 307

\bibitem[{{Wanderman} \& {Piran}(2010)}]{wp10}
{Wanderman}, D., \& {Piran}, T. 2010, MNRAS, 406, 1944

\bibitem[{{Zauderer} {et~al.}(2013){Zauderer}, {Berger}, {Margutti}, {Pooley},
  {Sari}, {Soderberg}, {Brunthaler}, \& {Bietenholz}}]{zbm+13}
{Zauderer}, B.~A., {Berger}, E., {Margutti}, R., {et~al.} 2013, ApJ, 767, 152

\bibitem[{{Zauderer} {et~al.}(2011){Zauderer}, {Berger}, {Soderberg}, {Loeb},
  {Narayan}, {Frail}, {Petitpas}, {Brunthaler}, {Chornock}, {Carpenter},
  {Pooley}, {Mooley}, {Kulkarni}, {Margutti}, {Fox}, {Nakar}, {Patel},
  {Volgenau}, {Culverhouse}, {Bietenholz}, {Rupen}, {Max-Moerbeck}, {Readhead},
  {Richards}, {Shepherd}, {Storm}, \& {Hull}}]{zbs+11}
{Zauderer}, B.~A., {Berger}, E., {Soderberg}, A.~M., {et~al.} 2011, Nature,
  476, 425

\end{thebibliography}

\end{document}